\documentclass[
    ,final            
  ]
  {aipproc}

\usepackage{graphics}
\usepackage{amsmath}
\usepackage{amssymb}

\layoutstyle{6x9}

\newcommand{\fR}{$f(R)$ }

\newcommand{\la}{\langle}
\newcommand{\ra}{\rangle}

\def\l{\left}
\def\r{\right}

\def\mat#1  { \begin{matrix}#1\end{matrix} }
\def\pmat#1 { \begin{pmatrix}#1\end{pmatrix} }
\def\cas#1  { \begin{cases}#1\end{cases} }

\def \D {\tilde{\nabla}}

\def\lgl{\langle}
\def\rgl{\rangle}

\newcommand{\curl}{\,\mbox{curl}\,}

\begin{document}

\title{CMB Tensor Anisotropies in Metric $f(R)$ Gravity}


\author{Hassan Bourhrous}{
  address={Astrophysics, Cosmology and Gravity Centre (ACGC), University of Cape Town, 7701 Rondebosch, Cape Town, South Africa.},
altaddress={ Department of Mathematics and Applied Mathematics, University of Cape Town, 7701 Rondebosch, Cape Town, South Africa.},
}

\author{\'Alvaro~de~la~Cruz-Dombriz}{
  address={Astrophysics, Cosmology and Gravity Centre (ACGC), University of Cape Town, 7701 Rondebosch, Cape Town, South Africa.},
altaddress={ Department of Mathematics and Applied Mathematics, University of Cape Town, 7701 Rondebosch, Cape Town, South Africa.},
}

\author{Peter Dunsby}{
  address={Astrophysics, Cosmology and Gravity Centre (ACGC), University of Cape Town, 7701 Rondebosch, Cape Town, South Africa.},
altaddress={ Department of Mathematics and Applied Mathematics, University of Cape Town, 7701 Rondebosch, Cape Town, South Africa.},
}

\begin{abstract}
We present a description of CMB anisotropies generated by tensor perturbations in $f(R)$ theories of gravity. The temperature power spectrum in the special case of $f(R)=R^n$ is computed using a modified version of CAMB package.
\end{abstract}

\maketitle

\section{Introduction}
The cosmic microwave background (CMB) carries information from the last scattering surface that puts constraints on the multitude of proposed cosmological models and the gravity theories they are based on. Amongst such theories are the $f(R)$ theories of gravity which have become an interesting endeavour to correct for the degeneracies of the concordance model \cite{Carloni:2004kp,delaCruzDombriz:2006fj,Carloni:2007yv,delaCruzDombriz:2008cp,Abebe:2011ry}.

\section{The Perturbation Equations}
%
In the $1+3$ formalism \cite{Ellis:1989jt}, spacetime is sliced into constant time hyper-surfaces with respect to fundamental observers with 4-velocity $u^a \equiv \frac{dx^a}{d\tau}$ where $\tau$ is the proper time along the observers' world lines. $u_a$ is time-like ($u^a u_a = -1$). The metric tensor can be decomposed into time and rest-space projection tensors: $h_{ab} = g_{ab} + u_a u_b$.
The derivatives in the $1+3$ split are defined as:
\begin{align}
& \mathrm{Projected\ covariant\ derivative:} && \D_a T_{b\dots}{}^{c\dots} \equiv h^d_{\phantom da} h^e_{\phantom eb} \dots h^c_{\phantom cf} \dots \nabla_d T_{e\dots}{}^{f\dots},\\
& \mathrm{Time\ derivative:} &&\dot T_{b\dots}{}^{c\dots} \equiv  u^a \nabla_a T_{b\dots}{}^{c\dots},\\
& \mathrm{Generalized\ 3D\ curl:} &&\curl T_{ab\dots c} \equiv u^n \eta_{dne(a} \D^d {T_{b\dots c)}}^e,
\end{align}
where $\eta_{abcd}$ is the totally antisymmetric tensor on spacetime. 

Kinematics are obtained by taking the irreducible decomposition of $\nabla_a u_b$:
\begin{equation}
\nabla_a u_b = \frac{1}{3} \Theta h_{ab} + \sigma_{ab} + \omega_{ab} - u_a A_b,
\label{velocity_decomp}
\end{equation} 
where $\Theta \equiv \D^a u_a = 3H$, $\sigma_{ab} = \sigma_{(ab)} \equiv \D_{\la a} u_{b \ra} = \l[ h_{(a}{}^c \, h_{b)}{}^{d} - \frac{1}{3} \, h_{ab} \, h^{cd} \r] \D_c u_d$, $\omega_{ab} = \omega_{[ab]} \equiv \D_{[a}u_{b]}$, and $A_b \equiv \dot u_b $ are the expansion scalar, the shear, the vorticity, and the acceleration respectively. $H$ is the local Hubble parameter. 
$A_a$, $\sigma_{ab}$, and $\omega_a$ characterize anisotropy in the expansion and therefore vanish in an exact FLRW universe \cite{Challinor:1999xz}.

The Riemann tensor, $R_{abcd}$, can be decomposed into its trace and trace-free parts \cite{Ellis:2009go}:
\begin{equation}
R_{ab}{}^{cd} = 2 g_{[a}{}^{[c} \, R_{b]}{}^{d]} - \frac{R}{3} \, g_{[a}{}^{[c} \, g_{b]}{}^{d]} + C_{ab}{}^{cd},
\end{equation}
where $C_{abcd}$ is the Weyl tensor and describes the vacuum contribution to the gravitational field. It can be decomposed further into the \emph{electric} and \emph{magnetic} Weyl tensors:
\begin{equation}
E_{ab}  \equiv  u^c u^d C_{acbd} \quad \mathrm{and} \quad H_{ab}  \equiv \frac{1}{2} \, \eta_{acde} \, u^e u^f C_{bf}{}^{cd}.
\end{equation}

The evolution of perturbations along the flow lines are described by seven propagation equations stated in \cite{Ananda:2007xh}, together with five spatial constraints. Here, we require only the linearised tensor mode equations characterized by: $\D^b E_{ab} = \D^b H_{ab} = \D^b \sigma_{ab} = \D^b \pi_{ab} = 0$, leaving us with five propagations equations and one spatial constraint, of which we state:
\begin{equation}
\dot \sigma_{ab} = - \frac{2}{3} \, \Theta \, \sigma_{ab} - E_{ab} + \frac{1}{2} \pi_{ab} \qquad \mathrm{and} \qquad H_{ab} = \curl \sigma_{ab}. \label{propagation_eq_tens}
\end{equation}
Equations \eqref{propagation_eq_tens} determine $E_{ab}$ and $H_{ab}$ from the shear. Further differentiation and tensor harmonic decomposition of the shear equation in \eqref{propagation_eq_tens} gives:
\begin{equation}
\ddot \sigma_k + \Theta \dot \sigma_k + \l[ \frac{k^2}{a^2} - \frac{1}{3} \l( \rho + 3 p \r) \r] \sigma_k = \frac{a}{k} \l[ \rho \dot \pi_k - \frac{1}{3} \l( \rho + 3 p \r) \Theta \pi_k \r].
\label{gwsk}
\end{equation}

\section{The Dynamics of \fR Cosmologies}
Metric variation of the \fR and matter actions \cite{Capozziello:2007ec}:
\begin{equation}
\label{actionfR} 
S = \int_\mathcal{V} f(R) \sqrt{-g} \text d^4x + S_{matter}
\end{equation}
leads to fourth order field equations (FE):
\begin{equation}
R_{ab} - \frac{1}{2} \ g_{ab} \ R  = \frac{1}{f_R} \ T_{ab} \ + \frac{1}{f_R} \l[ \frac{1}{2} \ g_{ab} \ f + \l(\nabla_a \nabla_b - g_{ab} \Box \r) f_R  - \frac{1}{2} \ f_R \ g_{ab} \ R \r],
\label{FE} 
\end{equation}
where $f_R \equiv \frac{\text df}{\text dR}$ and $T_{ab} \equiv -\frac{2}{\sqrt{-g}} \frac{\delta S_{matter}}{\delta g^{ab}}$ is the energy momentum tensor (EMT) of standard matter. From the latter, the density, pressure, energy flux, and anisotropic stress are isolated via: 
\begin{equation}
\rho = T_{ab}^{tot} u^a u^b,  \qquad 
p = \frac{1}{3} T_{ab}^{tot} h^{ab}, \qquad
q_a = -T_{cd}^{tot} u^c h^d{}_a, \qquad
\pi_{ab} = T_{\la ab \ra}^{tot}
\label{EMT_decomp_ops}
\end{equation}
%
%
\begin{equation} \mathrm{giving:\qquad }
\rho = \frac{\rho^m}{f_R} + \rho^R, \quad
p = \frac{p^m}{f_R} + p^R, \quad
q_a = \frac{q_{a}^m}{f_R} + q_{a}^R, \quad
\pi_{ab} = \frac{\pi_{ab}^m}{f_R}+\pi_{ab}^R.
\label{anis_decomp}
\end{equation}
The linearised form of $\rho^R$, $p^R$, $q^R_a$, and $\pi^R_{ab}$ are listed in \cite{Ananda:2007xh}. We give here the most important
contribution:

\begin{equation}
\pi^{R}_{ab} = \frac{1}{f_R} \l[ f_{RR} \D_{\lgl a} \D_{b\rgl}R - f_{RR} \, \sigma_{ab}\dot{R} \r] \label{curvature_anis}
\end{equation}

The $f(R)$ Friedmann equations in flat FLRW with vanishing cosmological constant:
\begin{equation}
\frac{1}{2} f - 3 f_R \frac{\ddot a}{a} + 3 \frac{\dot a}{a} \dot{R} f_{RR} = \rho_m \quad \mathrm{and} \quad \frac{1}{2} f - f_R (\dot{H}+3H^{2}) + \frac{1}{a} \frac{\text d}{\text dt} \l(a^2 \dot{R} f_{RR}\r) = - p_m
\label{friedmann_eqs}
\end{equation}

\section{CMB Tensor Anisotropies in $f(R) = R^n$ Gravity}
For $f(R)=R^n$ and a single fluid with constant equation of state (EoS), the scale factor is given by: $a(\eta) = \eta^\frac{2n}{3(1 + \omega) - 2n}$ \cite{Capozziello:2007ec}. In the radiation dominated era: $\pi^m \approx \pi^\gamma \approx 0$ and $\omega = \frac{1}{3}$. Separating the curvature and standard matter components, changing to conformal time, making the variable change $u_k = a^m \sigma_k$ and choosing $m = \frac{2-n}{n}$, Eq. \eqref{gwsk} becomes:
\begin{equation}
u_k'' + \l[k^2 - 2 \eta^{-2} \r] u_k = 0
\label{EoM}
\end{equation}
for all values of $n \neq 2$. This is the same equation of motion as for general relativity (GR).

In addition, Equation \eqref{curvature_anis} reduces to a direct relation between the shear and the curvature anisotropy:
\begin{equation}
\pi^{R}_k = - \frac{k}{a} \frac{1}{\rho} \frac{f_{RR}}{f_R} \dot{R} \, \sigma_k = - \frac{k}{a^2} \l( \frac{n-1}{\rho} \r) \frac{R'}{R}  \, \sigma_k
\label{curvature_anis_tensor_k_Rn}
\end{equation}
The evolution of the scale factor in $R^n$ gravity may be obtained from Eqs. \eqref{friedmann_eqs}.
%
\section{Results and Discussion}

\begin{figure}[htbp]
\includegraphics[width=0.58\textwidth]{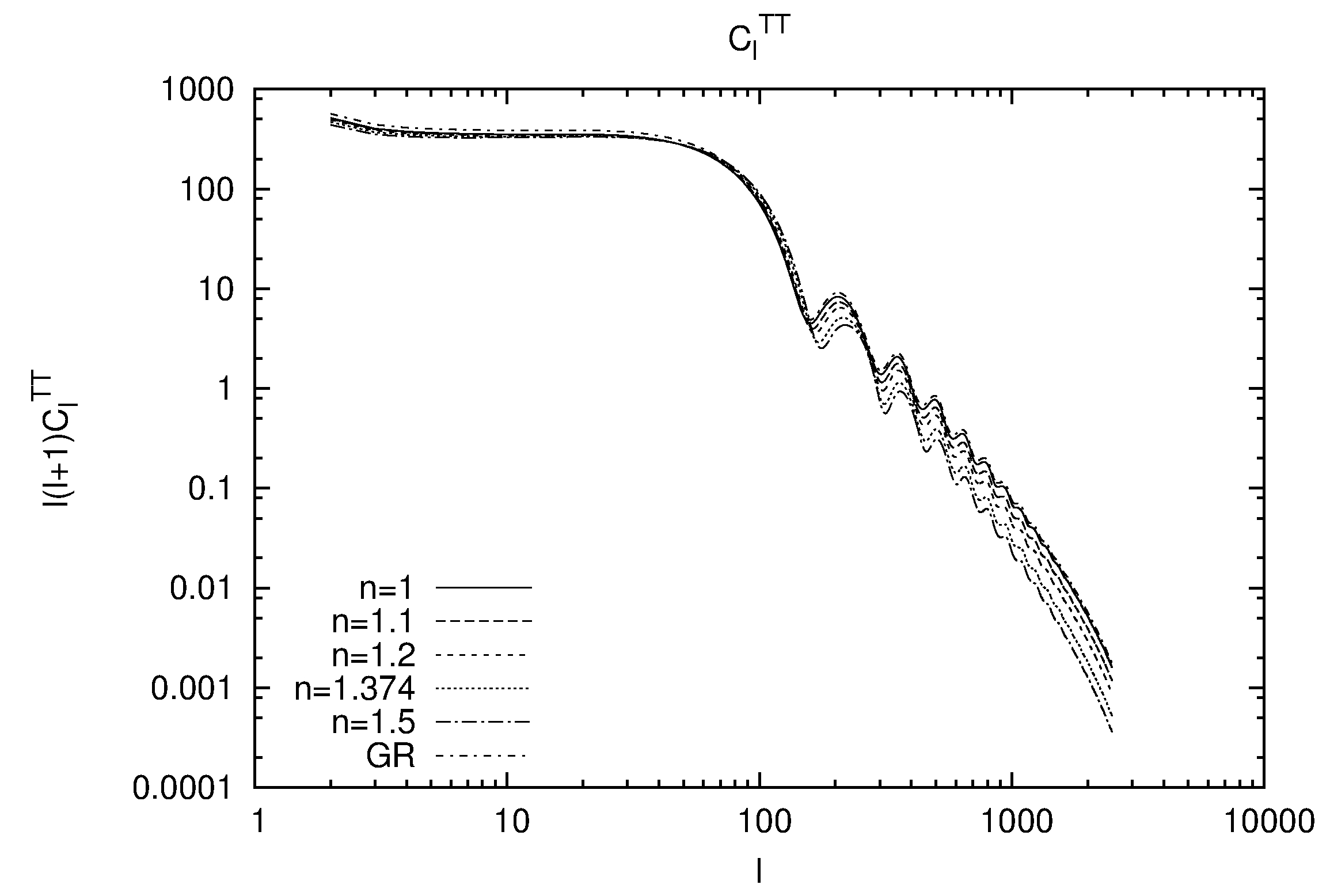}
\caption{\footnotesize The temperature power spectrum for tensor perturbations in $f(R) = R^n$ gravity ($n=1,\ 1.1,\ 1.2,\ 1.374,\ \text{and}\ 1.5$). The background is flat FLRW, $\Omega_b=0.035$, $\Omega_{CDM}=0.315$, $\Omega_{\Lambda}=0.65$, and  $\Omega_{\nu\text{-massive}}=0$. $H_0=70\, \text{km~s}^{-1} \text{Mpc}^{-1}$. No secondary anisotropies or reionization effects were considered.
} 
\label{ctt}
\end{figure}

Keeping a GR background allows to directly compare the influence of the first order modified equations. This is preliminarily our approach. Figure \eqref{ctt} shows the temperature power spectrum for $f(R)=R^n$, $n=1,\ 1.1,\ 1.2,\ 1.374,\ \text{and}\ 1.5$. For comparison, the power spectrum for $\Lambda$CDM is also plotted. It was obtained by running CAMB \cite{Lewis:2000PhD} on the same background but with the original GR evolution equations.

A few conclusions can be inferred directly from the power spectrum. We start by mentioning that the curve for GR and for $R^n = R$ are very close but not identical. The difference appeared after the use of the input file for the EoS. The plots for the two spectra are identical if a constant EoS for dark energy is used; typically $\omega = -1$. Also, one notices that the features of the power spectra are shifted more and more towards small scales with increasing power of $R$. The departure from GR increases with increasing wavenumber and power of $R$ as expected from the terms $k$ and $n-1$ in the numerator of the RHS of Eq. \eqref{curvature_anis_tensor_k_Rn}. The power difference between spectra for different values of $n$ becomes larger after multipole $\approx 160$. Finally, the power decreases with increasing powers of $R$ except for the interval $62 \lesssim l \lesssim 160$ in the $TT$ power spectrum where the opposite happens.

Calculation of the evolution of perturbations in $R^n$ gravity models with the exact cosmological scale factor is a work in progress. Equations \eqref{EoM} and \eqref{curvature_anis_tensor_k_Rn} together with Figure \eqref{ctt} are the main result of this work.


\begin{theacknowledgments}
\noindent HB would like to thank Dr. Garry Angus for useful discussions, the NRF/NASSP for financial support during his studies and the organisers of the Spanish Relativity Meeting (ERE2011) for a conference grant.
AdlCD acknowledges support from MICINN (Spain) project numbers FIS 2008-01323, FIS2011-23000, FPA 2008-00592, FPA2011-27853-01, Consolider-Ingenio MULTIDARK CSD2009-00064 (Spain) and URC (South Africa).
\end{theacknowledgments}



\bibliographystyle{aipproc}   

\bibliography{references}

\IfFileExists{\jobname.bbl}{}
 {\typeout{}
  \typeout{******************************************}
  \typeout{** Please run "bibtex \jobname" to optain}
  \typeout{** the bibliography and then re-run LaTeX}
  \typeout{** twice to fix the references!}
  \typeout{******************************************}
  \typeout{}
 }

\end{document}